*Review*

# A Contemporary Survey on 6G Wireless Networks: Potentials, Recent Advances, Technical Challenges and Future Trends

**Syed Agha Hassnain Mohsan** [1,*], **Yanlong Li** [1,2]

[1] Optical Communications Laboratory, Ocean College, Zhejiang University, Zheda Road 1, Zhoushan 316021, China; hassnainagha@zju.edu.cn (S.A.H.M.); lylong@zju.edu.cn (Y.L.)

[2] Ministry of Education Key Laboratory of Cognitive Radio and Information Processing, Guilin University of Electronic Technology, Guilin 541004, China

* Correspondence: hassnainagha@zju.edu.cn

**Abstract:** Smart services based on Internet of everything (IoE) are prophesied to reap notable attention by both academia and industry in the future. Although fifth-generation (5G) is a promising communication technology, however it cannot fulfill complete demands of novel applications. Sixth-generation (6G) technology is envisaged to overcome limitations of 5G technology. The vision and planning of future 6G network has been started with this aim to meet the stringent requirements of mobile communication. Our aim is to explore recent advances and potential challenges to enable 6G technology in this review. We have devised a taxonomy based on computing technologies, networking technologies, communication technologies, use cases, machine learning algorithms and key enabler technologies. In this regard, we subsequently highlight potential features and key areas of 6G. Key technological breakthroughs which include quantum communication, tactile communication, holographic communication, terahertz communication, visible light communication (VLC) Internet of Bio Nano Things, which can put profound impact on wireless communication, have been elaborated at length in this review. In this review, our prime focus is to discuss potential enabling technologies which can develop seamless and sustainable network, encompassing symbiotic radio, blockchain, new communication paradigm, VLC and terahertz. These transformative possibilities can drive the surge to manage the rapidly growing number of services and devices. In addition, we have investigated open research challenges which can hamper the performance of 6G network. Finally, we have outlined several practical considerations, 6G key projects and future directions. We envision 6G undergoing unprecedented breakthroughs to eliminate technical uncertainties and provide enlightening research directions for subsequent future studies. Although it is impossible to envisage complete details of 6G, we believe this study will pave the way for future research work.

**1. Introduction**

Commercial deployment of 5G has been started in 2019. It will mark a new era of a digital society and introduces innovative breakthroughs in terms of mobility, data rates, latency and communication [1]. If we look at the development of previous technologies, their subsequent utilization remains for almost 10 years. That is, the research of next generation starts with the commercialization phase of previous generation. As 5G has reached to its commercialization phase, it is the right time to launch next 6G. Some countries have already made strategic plans for 6G [2]. They have started 6G projects for timely deployment. In 2018, Finland introduced 6Genesis Flagship program with $290 million investment on 6G ecosystem [3]. German, South Korea and UK governments have invested in 6G quantum technology, while USA has also started projects on terahertz (THz) [4] 6G wireless networks. The Ministry of Industry and Information Technology of China has also focused on the development of 6G. The key technologies and novel services of 6G will mark a revolution in wireless networks. Japanese government has also started 6G projects [5].

The rapidly growing research on 6G and its emerging technologies with associated applications will marks a never-ending growth in this domain. International Telecommunication Union (ITU) has predicted up to 5 zettabytes global mobile data till 2030 [6] as shown in figure 1. Meanwhile, due to the emergence of the smart cities, e-health, smart industry and Internet-of-Everything (IoE) paradigm, there is an urgent need to focus on ultra-reliable low-latency communications (URLLC) which can enable a networked society. Besides offering massive data, the upsurge of IoE will support a myriad of new data services. Additionally, the promising IoE services entail integrating features like communication, computing and control into a single network architecture. In order to support the forefront services and meet their heterogeneous desiderata, several challenges must be address. These challenges include providing flexibility in the network architecture, monitoring the network performance, leveraging the sub-terahertz (THz) bands and designing a holistic orchestration strategy to integrate all network resource functionalities such as sensing, computing, control and communication in a scalable, efficient, intelligent and sustainable manner.

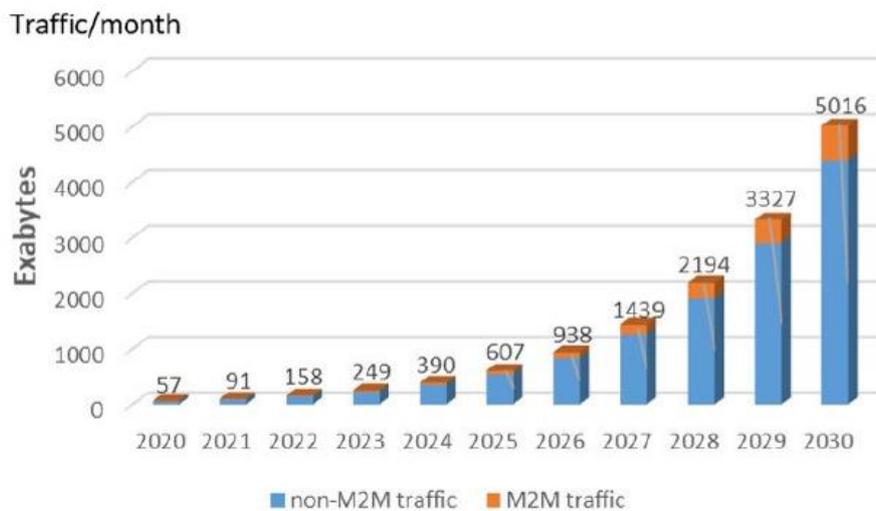

**Figure 1.** 2020-2030 Global mobile data traffic predicted by ITU [6].

Upcoming applications such as self-driving, smart city and e-health have stringent demands of throughput, data rate and latency which is beyond the limits of 5G. It is anticipated that 5G services will be widely available in a decade and then emerging 6G technology will pave its way to industry. The technical prospects of 6G include:
- Ultra-low latency and ultra-high data.
- Energy efficient resource-constrained devices.
- Ubiquitous network coverage.
- Intelligent and trusted connectivity.

6G will alter the perception and definition of communication, industry, society and modern lifestyle. 6G will revolutionize several technological domains which are yet to be envisioned. Besides its advantages, several critical problems exist in deploying 6G. In this article, we investigate and exploit these potential challenges in 6G. We have also analyzed and compared 5G, B5G and 6G.

The hype about 5G in media, industry and academia is highly validated by its prominent features with regards to data rate, reliability and accessibility of mobile services. Concretely, a paradigm shift in its design architecture has made 5G suitable to solve real business requirements [7].

The prominent features and promises of 6G technology have got attention from research fraternity. It is expected 6G will mark revolution in diverse domains from 2030 onward. Various aspects of next 6G are being considered in top-tier forums and several requirements of 6G have been collected [9-11]. S. Nayak et al. [8] have exposed 6G communication challenges. Moreover, several algorithms have been reported for 6G [12-13].

6G applications are vulnerable to some uncertainties. Autonomous systems and connected robotics depend on VLC and AI technology where data transmission, encryption and malicious behaviors can be intricate. The multisensory XR applications use quantum communication, terahertz communication and molecular communication technology, which make them susceptible to data transmission, access control risks and malicious behavior exposure. Wireless brain-computer interactions also utilize multisensory XR applications, but have own privacy and security problems. The main crucial weaknesses are encryption and malicious behavior. Although distributed ledger technologies and blockchain are secure, but they can also face malicious behavior. All inclusive, new areas of 6G are vulnerable to communication, encryption, malicious behavior, access control and authentication issues.

## 2. Our Survey

While several operators have announced plans to rollout 5G services, research in 6G technological trends has secured high impulse both in academia and industrial sectors. A number of research studies have reported key technological trends, potential issues and future research aspects which can bring 6G into reality, such as, see [14-15]. In [14-15], authors have provided a speculative study which addresses use cases, trends, technologies and briefly discussed associated challenges and future research aspects. In this article, we have adopted an approach to analyze the research challenges associated with 6G networks. We expect a combination of evolution of current networks and breakthrough technologies will be investigated in future. We also believe in our findings to promote research efforts towards promising technologies to meet the stringent demands of 6G.

An overview of 6G is illustrated in figure 2 which highlights key aspects in the form of localization, data rate, capacity and reliability in terms of energy per bit, jitter, latency, frame data rate [16]. In addition, overview of 1G to 6G with its applications is presented in figure 2 [17]. Whereas, we have also provided comparison between 5G and 6G parameters in Table I [16], [18].

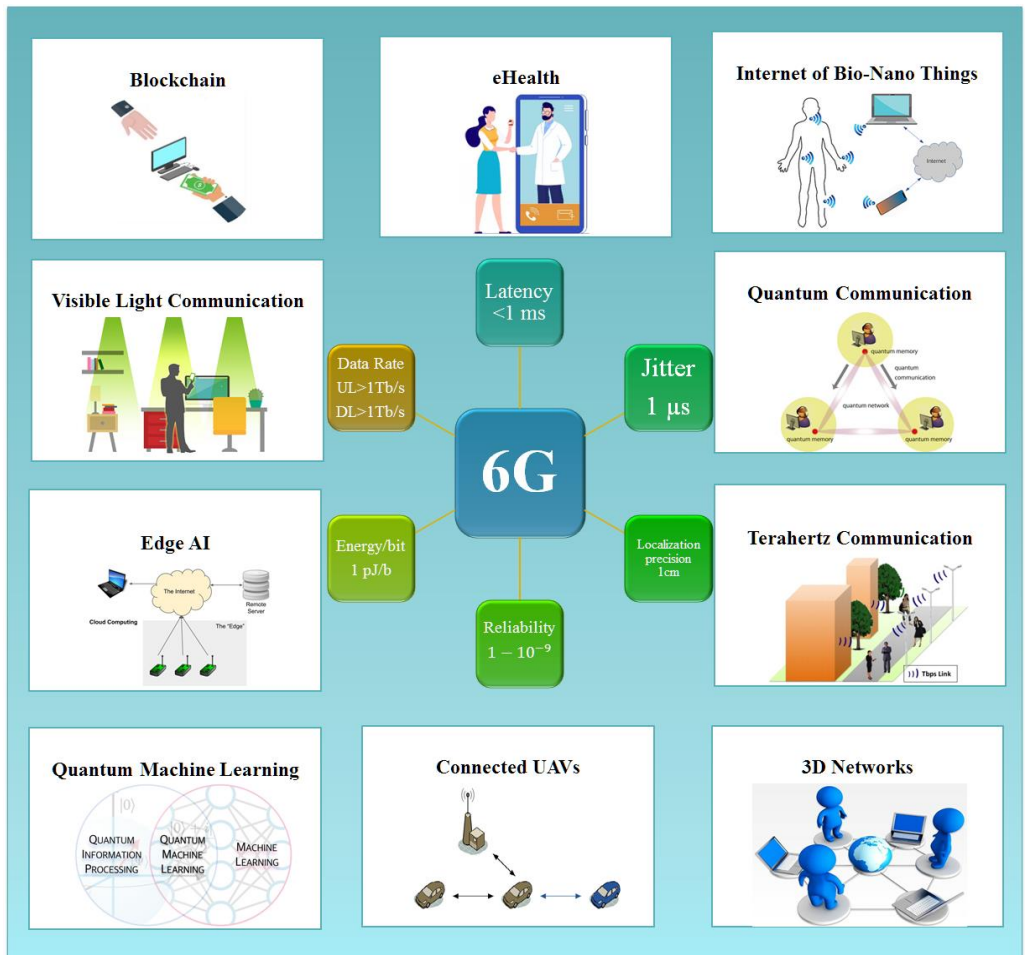

**Figure 2.** 6G wireless systems overview.

The studies discussed in [10], [15-16], [18-21], have focused on key enabler technologies of 6G. To the best of our knowledge, we are among few research groups who have provided taxonomy and state-of-the-art for 6G as given in Table I. Additionally, we have discussed potential challenges and future research directions. We have also suggested risk mitigation techniques in this article. Hence, our major contributions include:

1) A comprehensive overview of various 6G topics with highlighting recent academic activities and industry developments in different aspects of 6G.
2) The emerging key technologies are outlined with detailed explanation of potential issues.
3) An overview of 6G applications and future aspects are discussed.
4) The state-of-the-art towards 6G is provided.
5) A taxonomy based on machine learning techniques, communication technologies, computing technologies, use cases, key enablers and network technologies is provided.
6) Research challenges and associated solutions are discussed.
7) The privacy and security concerns are investigated and presented.
8) An outlook for future directions is provided.
9) For researchers, this review is devoted to open new horizons by guiding towards future research perspectives as it includes new references which can enable the pursuit of 6G.

The rest of our survey is organized as follows. Evolution of mobile communication networks is presented in section III. We have briefly discussed current research towards 6G in section IV. Section

V outlines the state-of-the-art advances toward enabling 6G wireless networks. Section VI the devised taxonomy. Key areas in 6G networks are listed in section VII. Section VIII presents vision and key features for 6G. Potential challenges and applications are discussed in Section IX and X respectively. Finally, we have concluded this study in Section XI.

**TABLE 1.** COMPARISON SUMMARY OF THE EXISTING SURVEYS

| Reference | Use cases | Key enablers | Taxonomy | Recent advances |
|---|---|---|---|---|
| Giordani et al. [10] | Yes | Yes | No | No |
| Saad et al. [15] | Yes | Yes | No | No |
| Chen et al. [18] | No | Yes | No | No |
| Letaief et al. [19] | Yes | Yes | No | No |
| Akyildiz et al. [20] | Yes | Yes | No | No |
| Kato et al. [21] | No | No | No | No |
| Yang et al. [22] | No | Yes | Yes | Yes |
| Zhang et al. [23] | Yes | Yes | No | No |
| Khan et al. [24] | Yes | Yes | Yes | Yes |
| Tariq et al. [25] | No | No | Yes | Yes |
| Zong et al. [26] | No | Yes | Yes | Yes |
| Our Survey | Yes | Yes | Yes | Yes |

## 3. Evolution of Mobile Communication Network

A phenomenal advancement is witnessed in mobile communication networks sincere the emergence of first generation in 1980s. This advancement contains several generations having different techniques, technologies, data rate, capacities and standards. Every new generation is introduced in almost a time span of 10 years [27]. Figure 3 presents the evolution of mobile networks.

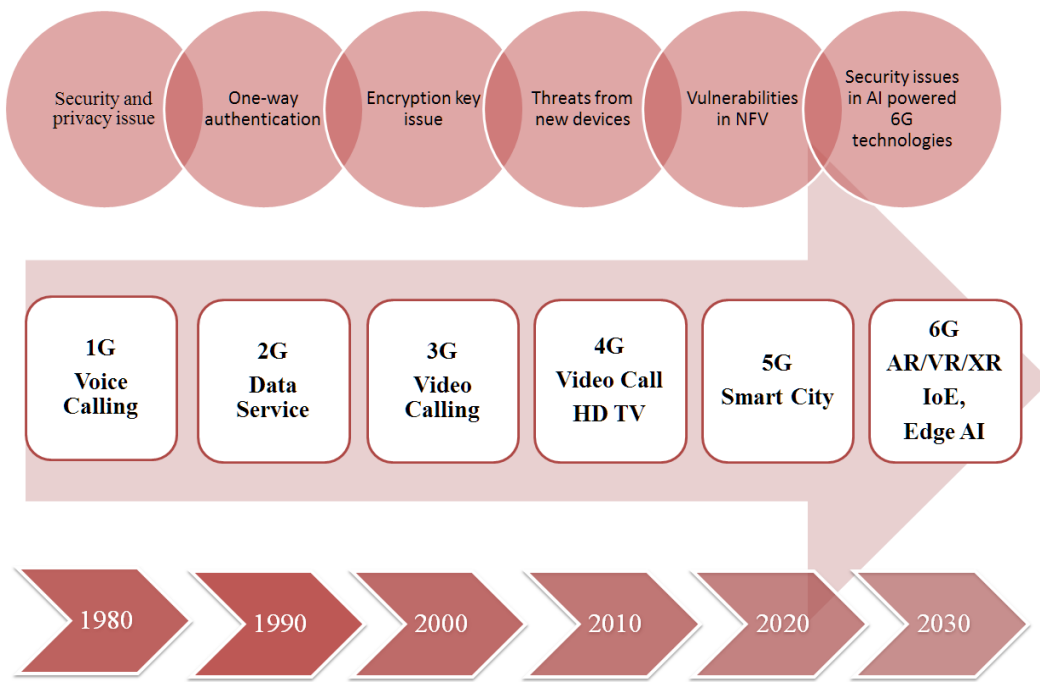

**Figure 3.** Evolution of wireless mobile technologies

*3.1. From 1G to 3G*

1G was developed in 1980 from voice calling with 2.4 kbps data rate. Data transmission was made in the form of analogue signal without any universal wireless standard. It led to several drawbacks e.g. security issues, low transmission efficiency and hand-off [28]. In 1990, 2G was introduced and it was dependent on digital modulation techniques e.g. Code Division Multiple Access (CDMA) and Time Division Multiple Access (TDMA). It supported data rate of 64 kbps featuring both Short Message Service (SMS) and better voice calling. Global System for Mobile Communication (GSM) was the dominant standardization in 2G era [29]. In 2000, 3G was introduced with the aim to transmit data at high-speed. 3G network provides high speed access to internet and 2 Mbps data transfer rate [30]. It covers advanced features as compared to 1G and 2G, including Video services, navigational maps, live streaming and web browsing. To support global coverage, Third Generation Partnership Project (3GPP) was developed to find technological aspects and standardizations [31].

*3.2. 4G*

4G was introduced in 2000s. It is IP based network which can feature data rate up to 1 Gbps for downlink communication and 500 Mbps for uplink communication respectively. Apparently it can reduce latency and enhance spectral efficiency. It is capable to meet required criteria set by video chatting, HD TV content and Digital Video Broadcasting (DVB). In addition, it provides automatic roaming to facilitate wireless service anywhere and at any time.

*3.3. 5G*

5G has completed its initial testing, standardization processes and paved its way to commercialization in few countries. China, UK, South Korea and USA have launched 5G technology [32]. The main target of 5G is to revolutionize energy efficiency, network reliability, latency, data rates and massive connectivity [33]. It makes use of both mmWave and microwave bands to enhance data transmission up to 10 Gbps. 5G features access technologies like Filter Bank multi carrier (FBMC) and Beam Division Multiple Access (BDMA). Some emerging technologies e.g. software Defined Networks (SDN), Massive MIMO, Information Centric Networking (ICN) and network slicing are also integrated into 5G [34-36]. IMT 2020 has suggested three key usage scenarios: Massive machine type communications (mMTC), Enhanced mobile broadband (eMBB) and Ultra-reliable and low latency communications (URLLC).

*3.4. Vision of Green 6G*

As 5G has entered into commercialization phase, research fraternity around the globe has started focusing on future 6G technology which is expected to be launched in 2030s. This progress of 5G yields the conceptualization of 6G with the capability to unleash the promises of ample autonomous services. Specifically, 6G is envisaged to offer innovative promising wireless techniques and novel network designs into perspective. 6G can bring a remarkable advancement in wireless technology with ultra-low latency in microseconds and data rates up to 1 Tbps. Its capacity is envisioned to be 1000 times higher than 5G with spatial multiplexing and THz frequency communication. One main objective of 6G is to feature ubiquitous coverage by incorporating undersea communication and satellite communication to support global coverage [37]. Haptics communication, quantum machine learning and energy harvesting technologies will put profound impact to realize future sustainable green networks. More precisely, it has the capability for high-precision communications for tactile services to enable the desired sensing experience at various steps, such as smell, touch, vision and listening. 6G is defined by its three classes as: ultrahigh data density (uHDD), ultrahigh-speed-with low-Latency communications (uHSLLC) and ubiquitous mobile ultra-broadband (uMUB). Table II illustrates a comparison between 5G and 6G while Table III summarizes evolution from 1G to 6G. 6G is expected to fill gap of radio coverage limitation in previous generations. We can say it will accommodate the whole surface area of ear including airspace, forests, deserts and oceans as complete vision of 6G can be seen in figure 4.

The main technical aspects to realize this vision of 6G include:

- To meet the extreme high level of communication reliability.
- Offering ultra-high throughput and high data rates to support massive connectivity even in extreme conditions.
- Providing the required quality of immersion and unified quality of experience required by extended reality (XR) applications.
- Delivering real-time tactile feedback to meet the targeted haptic applications like digital healthcare.
- Integrating AI to enable seamless connectivity to control environments like smart city, smart industry, self-driving system and smart structure.

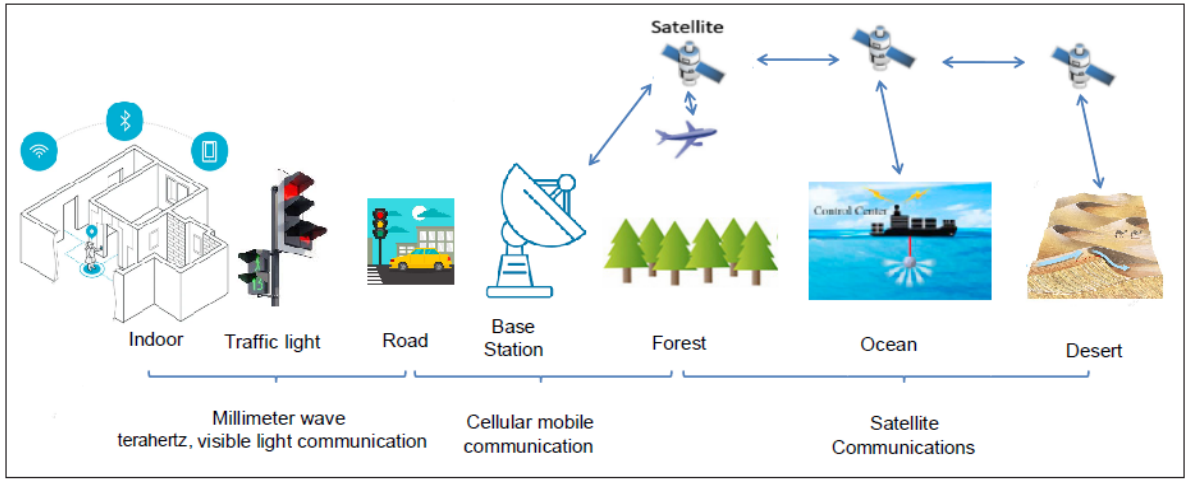

**Figure 4.** Vision of future 6G technology

TABLE II. EVOLUTION FROM 5G TO 6G

| Key parameters or characteristics | 5G | 6G |
|---|---|---|
| Reliability | $10^{-5}$ | $10^{-9}$ |
| Mobility (km/h) | 350-500 | 1000 |
| End-to-end latency (ms) | 1 | 0.1 |
| Area traffic capacity (Mbps/m$^2$) | 10 | 1000 |
| Energy Efficiency (Tb/J) | NA | 1 |
| Spectral efficiency (b/s/Hz) | 0.3 | 3 |
| Peak spectral efficiency (b/s/Hz) | 30 | 60 |
| Connection Density (device/km$^2$) | $10^6$ | $10^7$ |
| User Data rate (Gbps) | 1Gb/s | >10Gb/s |
| Peak data rate | 10-20Gb/s | >100Gb/s |
| Channel bandwidth (GHz) | 1 | 100 |
| Receiver sensitivity | -120dBm | <-130dBm |
| Coverage | 70% | >99% |
| Position precision | m | cm |
| Localization precision | 10 cm on 2D | 1 cm on 3D |
| Delay | ms | <ms |
| Processing delay | 100 ns | 10 ns |
| Jitter | 1 μs | 0.1 μs |
| Automatic integration | Partial | Full |
| Haptics communication | Partial | Full |
| THz communication | No | Yes |
| XR/AI integration | Partial | Full |
| Intelligent Reflecting Surface (IRS) | Conceivable | Yes |
| Satellite integration | No | Full |
| Cell free networks | Possible | Yes |
| Real-time buffering | No | Yes |
| Pervasive AI | No | Yes |
| VLC | No | Yes |
| Center of gravity | User | Service |
| Technique | m-MIMO | SM-MIMO, UM-MIMO |
| Energy consumption | Low | Ultra-low |
| Device lifetime | 10 years | 40 years |
| Dependability | Not considered | Relevant |
| End-to-End optimization | Not considered | Relevant |
| Device type | Sensors, smartphones and drones | Distributed Ledger Technology (DLT), Smart implants, Connected Robotic and Autonomous System (CRAS) |
| Services | eMBB, URLLC, mMTC | HCS, MPS, MBRLLC, mURLLC |

TABLE III. 1G-6G TECHNOLOGIES CHARACTERISTICS

| Feature | 1G | 2G | 3G | 4G | 5G | 6G |
|---|---|---|---|---|---|---|
| Time span | 1980-1990 | 1990-2000 | 2000-2010 | 2010-2020 | 2020-2030 | 2030-2040 |
| Highlight | Mobile | Digital format | Internet connectivity | Real-time applications | Extreme data rates | Secrecy, privacy, security |
| Core network | PSTN | PSTN | Packet N/W | Internet | IoT | IoE |
| Utility | Voice calling | SMS | Image | Telecasting | 3D VR/AR | Quantum |
| Frame work | SISO | SISO | SISO | MIMO | Massive MIMO | Intelligent Surface |
| Frequency band | 800 MHz | 890-960 MHz | 1.94-2.14 GHz | 0.3-3 THz | 30-300 GHz | 0.3-3 THz |
| Maximum Data rate | 2.4 kb/s | 144 kb/s | 2 Mb/s | 1 Gb/s | 35.46 Gb/s | 100 Gb/s |
| Transmission Range | - | 35 km | 10 km | 5 km | Below 1 km | Below 1 km |
| Multiplexing | FDMA | FDMA, TDMA | CDMA | OFDMA | OFDMA | Smart OFDMA plus IM |
| Application | Voice calling | Macro calling | Macro cell | Macro cell | Pico cell | Small cell |

**4. Current Research Progress towards 6G**

Several researchers have shown vision for 6G and many research institutes have started planning activities [38-40]. Referring to 6G vision, David et al. [41] suggested that service classes and battery lifetime of mobile device need special attention than latency and data rates. Raghavan et al. [42] pointed out 6G research should consider device manufacturing capability to design a closed loop of research plans. Yastrebova et al. [43] predicted new communication aspects including tactile internet, self-driving, UAVs and holographic connectivity. Tactile internet (TI) is an emerging paradigm envisaged to catalyze the development of a plethora of new services such as education, eHealth and smart manufacturing. To fully perceive TI, the communication infrastructure (CI) should meet stringent design requirements for TI. Particularly, the CI should enable high reliability and extremely low latency. Furthermore, it must fortify data privacy and security without imperiling the latency requirements. To meet these targeted desiderata and address new services with distinctive features, the maturing of disruptive 6G wireless communication technologies is of paramount significance.

It is expected future wireless communications will have a similar reliability as wires communications. Future trends and driving applications are discussed in references 38 and 39. In future, blockchain technology will offer satisfactory performance and simplify network controllability. Tariq et al. [25] have proposed human-centric service and key performance indicators along with proving a comprehensive comparison between 5G and 6G. Some recent articles have discussed practical scenarios including 6G data centers [44], intelligent reflecting surfaces (IRSs) [45] and multiple accesses [46]. Intelligent reflecting surface (IRS) is observed as an energy efficient technology to enlarge coverage area in future wireless technologies at low complexity and implementation cost. Networking patterns such as 3D super-connectivity, decentralized resource allocation and cell-less architecture are outlined in some studies [47-48]. Mahmood et al. [49] have elaborated vertical-specific wireless network and Machine-type communications (MTCs) which can provide unified solution to enable seamless connectivity in vertical industries.

Reconfigurable intelligent surfaces, artificial intelligence (AI) and terahertz communications are attractive technological aspects pertaining to 6G. Rappaport et al. [50] provided a comprehensive study of THz communications with practical demonstrations. Stoica et al. suggested that AI-integrated 6G can empower new features such as opportunistic set-up, self-configuration, context awareness and self-aggregation [51]. Moreover, AI-empowered 6G will enable a paradigm shifting perspective in mobile networks [52]. Quantum Machine learning algorithm for AI-empowered 6G is discussed in an article [53]. Renzo et al. have envisaged reconfigurable intelligent surfaces to lay hardware foundation of AI [54]. Reconfigurable intelligent surfaces are proposed for massive MIMO in some earlier studies [55-57]. Here we have presented some standardization efforts and research activities. A summary of research studies over 6G is provided in Table IV.

TABLE IV. SUMMARY OF RECENT RESEARCH STUDIES ON 6G

| Reference | Year | Research contributions | Key focus |
|---|---|---|---|
| Katz et al. [3] | 2018 | This study sheds some light on initial research of 6G technology and 6Genesis Flagship Program (6GFP). This article includes motivation, trends and future aspects of 6G. | 6Genesis Flagship Program (6GFP) |
| Letaief et al. [19] | 2019 | AI based 6G key technologies and applications are discussed. This study presents key trends in the evolution to 6G. | Artificial intelligence |
| Yang et al. [22] | 2019 | This article includes an overview of 6G promising techniques and key requirements. This study highlights potential challenges, solutions and security approaches. | 6G vision and potential techniques |
| Zhao et al. [45] | 2019 | This article outlines 6G challenges, future directions and a possible roadmap for AI based cellular networks. | MIMO and intelligent reflecting surfaces |
| Rappaport et al. [50] | 2019 | This article presents novel approaches, promising discoveries, key technologies and potential challenges for 6G. It discusses current standard body regulations for applications using above 100 GHz. It provides in-depth details of THz products and applications. | Challenges and opportunities of 6G |
| Stoica et al. [52] | 2019 | This study outlines AI revolution for future 6G networks. | AI |
| Nawaz et al. [64] | 2019 | A comprehensive study is provided in ML, QC and QML in order to seek challenging issues and potential benefits. A new QC-aided and QML-enabled framework for future technology is presented to articulate its enabling technologies and potential challenges at the user end, air interface, network edge and network infrastructure. Finally, this research study identifies some groundbreaking future research directions for B5G networks. | Quantum machine learning |
| Dang et al. [4] | 2020 | This study presents a systematic framework of 6G applications. It highlights communication technologies key potential features of 6G. Authors have investigated potential issues which can hamper deployment of 6G. | Secrecy, security and privacy |
| Giordani et al. [10] | 2020 | Authors have discussed technologies which will develop wireless networks towards 6G. They have presented key enablers, use cases and a full stack overview of 6G requirements. | 6G use cases and technologies |
| Saad et al. [15] | 2020 | A holistic vision of 6G technology is presented in this article. Primary drivers for 6G systems are identified including | 6G performance components |

| | | technological trends and applications. Authors have proposed a new set of service classes. A comprehensive research agenda and solid recommendations for the 6G roadmap is outlined in this research study. | |
|---|---|---|---|
| Gui et al. [11] | 2020 | This article outlines 6G core services, eight KPIs and two centricities. Authors have presented 6G architecture and outlined potential challenges, possible solutions and four application scenarios. | 6G key performance indices (KPIs) and core services |
| Mao et al. [13] | 2020 | This study proposes AI enabled adaptive security strategy for IoT networks in 6G technology. In this security method, IoT devices are linked to cellular networks through mmWave and THz. Authors have used EKF for efficient energy harvesting in 6G to avoid energy exhaustion. | QoS and security for 6G |
| Kato et al. [21] | 2020 | In this article, authors have analyzed machine learning techniques for 6G and highlighted 10 crucial challenges for advancing ML in 6G. | Challenges in machine learning for 6G |
| This review study | | In this study, we devise a taxonomy based on computing technologies, networking technologies, communication technologies, use cases, machine learning algorithms and key enabler technologies. We have briefly discussed 6G key projects, potential challenges and applications. | 6G technologies, key enablers, key areas, use cases, key projects, potential challenges and applications |

Apart from above discussions, some countries around the globe have started 6G projects to reshape the framework of 6G networks. In 2019, University of Oulu Finland started 6Genesis Flagship Program [58]. In March 2019, 6G research race was triggered in first 6G Wireless Summit organized in Levi, Finland. Many seminars and workshops have been conducted worldwide such as Carleton 6G, Wi-UAV Globecom 2018 workshop and Huawei 6G Workshop which was organized as a virtual event in March 2020 [59]. Beyond academia, 6G has also attracted governments, industrial organizations and standardizing bodies. In 2018, "Enabling 5G and beyond" was launched by IEEE. Google has launched Loon Project [60] to provide internet connectivity to five billion users from remote communities. In the end of 2018, Ministry of Industry and Information Technology, China made an official announcement to expand 6G research and investment. Korea Advanced Institute of Science and Technology (KAIST) has collaborated with LG Electronics to establish a 6G research center. SK Telecom, Ericsson and Nokia are collaborating in 6G research. The Federal Communications Commission (FCC), USA has opened 95 GHz -3 THz spectrum for research contributions on 6G. Moreover, Networking Research beyond 5G'project has been launched in Japan to use 100 GHz to 450 GHz THz spectrum. Different countries around the world such as Germany, Australia, and Sweden etc. are carrying out research on 6G. We have summarized country wise 6G initiatives in Table V.

TABLE V. 6G PROJECTS IN DIFFERENT COUNTRIES

| Country | Year | Research Initiative |
|---|---|---|
| 2018 | Finland | University of Oulu launched 6G initiative in 2018. UROS and University of Oulu has announced strategic partnership. University of Oulu has required Toyota self-driving car for research purposes. |
| 2019 | China | 37 research institutes have collaborated for 6G research. They have launched National 6G Technology Research and Development Promotion Working Group. |
| 2019 | USA | USA opened spectrum between 95 GHz and 3 THz. BWN Lab in Georgia Institute of Technology is working on 6G research projects. |
| 2019 | South Korea | KAIST has collaborated with LE Electronics to establish a 6G research center. |
| 2019 | Germany and France | German and French ministries have officially announced to develop 6G combat aircraft in order to bring revolution in military affairs. TU Berlin has established new Einstein fellowship to strengthen research in 6G. |
| 2020 | Japan | NTT, Sony and Intel have started collaboration for research on 6G technology. Japan has also made plans to invest $US 2 billion to carry out industrial research on 6G. |
| 2020 | Saudi Arabia | Research groups from KAUST have initiated 6G research. |
| 2020 | Brazil | 6G Brazilian Project was introduced to develop a national-wise framework for 6G networks. |
| 2021-2026 | South Korea | Government of Korea has planned to spend $169 million to secure 6G and it will start 6G pilot project around 2026. |

## 5. 6G: State-of-the-art

In this section, we present state-of-the-art approaches to enable 6G. Federated learning for edge network including Stackelberg-game-based incentive mechanism, hardware-software co-design and resource optimization is discussed in [61]. Finally, they have discussed potential challenges and future research plans. In-Edge AI provided good results for edge computing and caching. A 3D wireless cellular network using drones is demonstrated in ref. [62]. They provided an analytical approach for frequency planning and truncated octahedron cells for lowest number of drone base station. They considered two issues of network planning and 3D cell association in this article. An illustration of opportunities and critical challenges in THz communication is presented in ref. [63]. In this article, Mumtaz et al. investigated different standardization activities and available bands for THz communication. However, it is important to highlight key standards for 6G to incorporate THz range at this stage. Nawaz et al. [64] presented quantum machine learning in the context of 6G. They outlined state-of-the-art machine learning techniques, quantum communication schemes and investigated potential research challenges to implement quantum machine learning techniques in 6G. In [65], Salem et al. demonstrated an EM based model for blood through effective medium theory. They discussed advantages of proposed model for healthcare applications. S. Canovas-Carrasco et al. [66] developed architecture via THz communication for nano-networks. X. Wang et al. [67] proposed machine learning based In-Edge AI to empower intelligent edge computing. Double deep Q-learning network (DDQN), federated learning-based DDQN and Centralized DDQN have been proposed in this article. They designed two devices: nanorouters and nanonodes. They were able to carry out THz communication between nanonodes. They mitigated path loss and molecular absorption noise. In addition, they enhance transmission rate through energy harvesting by blood flow and an additional external source.   Basar et al. demonstrated that these intelligent surfaces can enhance the spectral efficiency [68] of 6G network.

**6. Taxonomy**

We consider communication technologies, computing technologies, machine learning schemes and key enablers to devise taxonomy as it can be seen in figure 5. Further details are provided in below subsections.

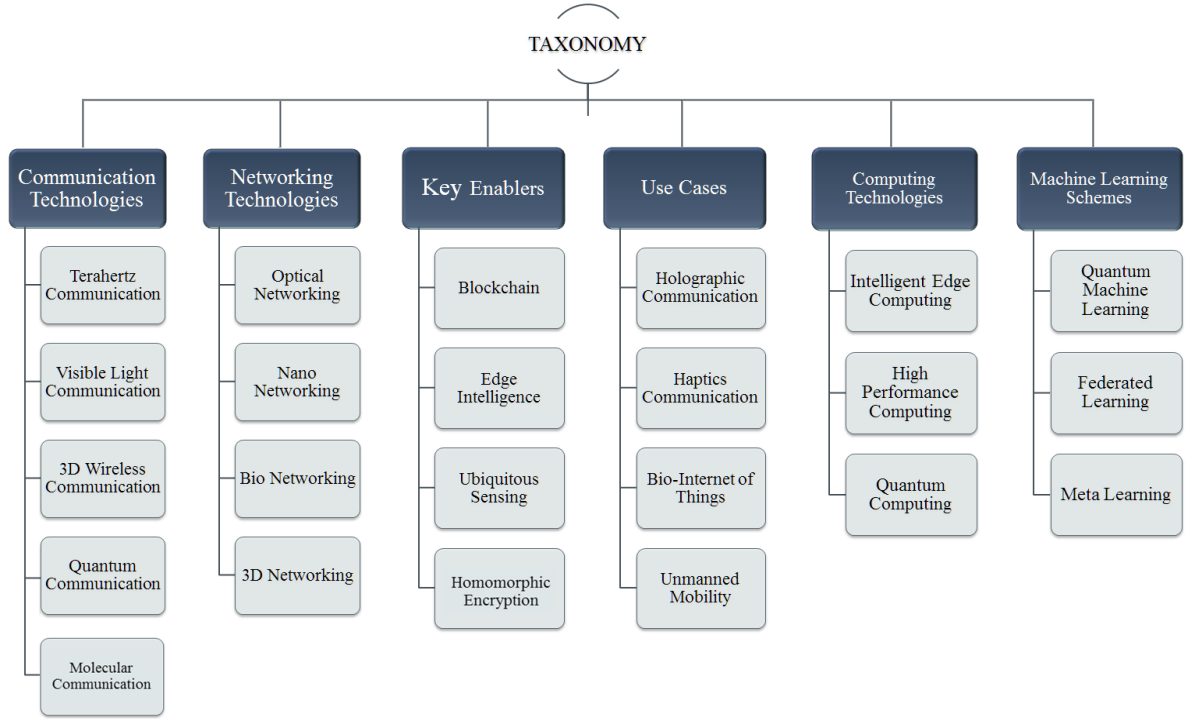

**Figure 5.** Taxonomy of 6G wireless systems.

*6.1. Communication Technologies*

*6.1.1. Terahertz Communication*

One key solution towards existing spectrum crunch is to utilize the THz-band, which is expected to assist the infrared (IR) and mmWave band, by offering a considerably wider bandwidth and supporting promising services with higher data rates requirements. It operates in the region of 100 GHz to 10 THz as shown in figure 6. It enables the potential of high data rates and high frequency connectivity. The main issues prevented to use THz in commercialization are high penetration loss, molecular absorption, propagation loss, RF circuitry and engineering challenges for antenna. In addition, THz communication can be improved by selecting frequency bands which are less affected by molecular absorption. THz communication is characterized by high security, moderate energy consumption, short range and robust to atmospheric conditions [69-71]. In fact low frequency channel model cannot capture the full characteristics of high frequency THz communication which experiences high molecular absorption and attenuation. Therefore, it is important to design realistic channel models for THz links to address LOS path in the THz communication system to investigate the performance limitations for such technology. On the other hand, THz communication needs to rethink current solutions and find new approaches which provide a seamless functionality over the complete THz band. Such as, designing efficient beamforming and tracking methods which can precisely and dynamically trace the location of THz-assisted devices is an open research problem. Additionally, there is need for research intervention to design tunable and intelligent ultra-fast modulators to support reliable and efficient THz communication links. Other open research issues in THz communication include novel hardware architecture designs and incorporation of massive MIMO and intelligent surfaces.

A dramatic increase in data traffic is witnessed recently. This exponential growth has put demand for better coverage and higher data rates [72]. THz (0.1-10 THz) communication is envisaged to be among key enabling technologies for future 6G. THz band can facilitate with ultra-fast massive data transfer to support plethora of applications. Federal Communications Commission (FCC) has issued frequency band above 95 GHz [73] for future contributions. Researchers should pay attention to multiple factor such as interference, imperfection in circuit and high complexity in realistic communication channel to enhance data rates. Although THz bands are used in object detection, imaging and radio spectroscopy, however it still needs research attention in wireless communication domain. THz band lies between IR and mmWave spectrum as shown in figure 6 while previously it was names as "no-man's land"'. Recently, a significant research progress has been made to realize wireless network on chip (WNoC) in THz [74]. Z. Chen et al. [75] have provided a comprehensive survey over THz communications.

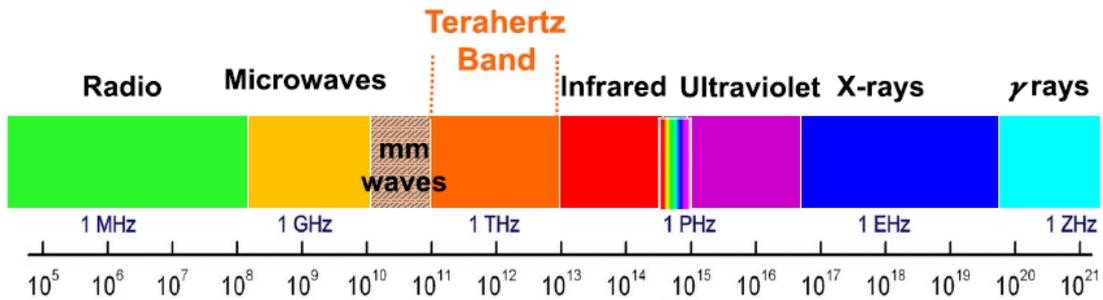

**Figure 6.** THz spectrum [20].

*6.1.2. Visible Light Communication (VLC)*

6G will support high coverage by incorporating undersea networks and space networks with terrestrial networks. As undersea and space/air networks vary from typical terrestrial network; therefore, typical EM waves are unable to attain high speed data for these environments. Optical communication utilizing laser diode can operate in these environments to achieve high speed data transmission. Meanwhile, visible light communication (VLC), operating between 430-790 THz [76], is a promising alternative to RF for future 6G. Since, VLC is functional in THz range thus it provides substantial bandwidth to meet the data rate and capacity needs of 6G. Taking 6G into account, a hybrid network can be designed to leverage the best of VLC and other optical or RF systems such as WiFi and Bluetooth (BLE).

VLC can be performed by using light-emitting diodes (LEDs). VLC receiver can be a photodetector or solar cell. It can be used for several application including indoor positioning, energy harvesting, diver-to-diver communication, vehicle-to-vehicle communication and underwater networks [77]. VLC offers inherit benefits including high data rates, safety, low cost deployment, robustness against interference, high energy efficiency and ultra-wide frequency band. VLC can be employed to future 6G applications. The main characteristics of VLC are communication and lightning. As compared to RF, VLC systems are considered intrinsically secure. This technology has been successfully used for an extensive range of applications including underwater mines, visible light identification system, underwater communication and vehicular communication. However, VLC will face several challenges like coverage, mobility, intercell interference and LED connectivity to internet [77]. Specifically, because of broadcast feature of VLC systems, VLC systems are vulnerable to eavesdropping threats at public places. The functionalities of VLC systems are different from RF systems which must be taken into account to develop PLS strategies. For example, VLC channels are real-valued and quasi-static. Therefore, such functional constraints must be reconsidered for the optimization and performance analysis of physical layer security (PLS) strategies in VLC systems. It is also important to mitigate mobility issue for seamless connectivity.

*6.1.3. 3D Communication*

3D communication is another leading aspect of 6G which integrates airborne and ground networks. In 3D communication, low-orbit satellites and unmanned aerial vehicles (UAVs) can be used as base stations (BSs) [78]. In comparison with 2D, 3D communication has significantly divergent nature due to altitude dimension. Thus, novel techniques are required to handle mobility and resource allocation.

*6.1.4. Molecular Communication*

Advance nanotechnology enables manufacturing of biosensors, implantable chips and nano-robots. It has various applications such as biomedicine and nanoscale sensing [79]. Specifically, biomedicine application can enhance health care through monitoring of body organs and intelligent drug delivery. Establishing connection between nanodevices and internet can transfer information and maintain effective communication. Internet of Bio-Nano-Things (IoBNT) can connect biological entities and nanodevices [79]. In addition, combining body area networks and IoBNT can provide feasible solution to enhance health care. This technique makes use of shorter wavelengths to communicate at 1 cm or m. The key challenges in this technique are channel modeling and transceiver design.

*6.1.5 Quantum Communication*

Another merging technology is quantum communication which will provide considerable security, long distance communication and higher data rate in 6G network [80-81]. It is a technique to deliver a quantum state from a sending component to the receiving component. It can execute the tasks which cannot be performed through classical techniques. Some of the appealing contributions of quantum communication are quantum network, Quantum Key Distribution (QKD), quantum teleportation, Quantum Secret Sharing (QSS) and Quantum Secure Direct Communication (QSDC). The high security mechanism of quantum communication makes is appropriate technology for future 6G. Particularly, the prime motive of quantum entanglement and its non-cloning theorem, inalienable law, superposition and non-locality offer strong privacy and security. The next generation of applications enabled by quantum communication are brain-computer interaction (BCI), tactile internet and intelligent communications. As it is contradictory to achieve both high data rate and long distance communication [82], new repeaters can be designed to achieve high data rate and secure long distance communications. Some research groups have already started working on quantum key distribution (QKD) and protocols. UAVs, high altitude stations and satellites can be selected as key redistribution or regeneration and nodes. Single photon emitter device can operate as quantum device above absolute zero temperature. A summary of existing research surveys is given in Table VI.

TABLE VI. SUMMARY OF THE EXISTING SURVEYS

| Technology | Reference | Security and privacy challenges |
|---|---|---|
| Artificial intelligence | [83] | Malicious threat |
| Artificial intelligence | [84] | Communication |
| Artificial intelligence and quantum communication | [85] | Encryption |
| AI | [86] | Access control |
| AI | [87] | Authentication |
| Blockchain | [88] | Communication |
| Blockchain | [89] | Access control |
| Blockchain | [90] | Authentication |
| Visible light communication | [91] | Malicious threat |
| Visible light communication | [92] | Communication |
| Terahertz communication | [93] | Malicious threat |
| Terahertz communication | [94] | Authentication |
| Quantum communication | [95] | Encryption |
| Molecular communication | [96] | Authentication |
| Molecular communication | [97] | Encryption |
| Molecular communication | [98] | Malicious threat |

*6.2. Networking Technologies*

Innovative networking technologies for 6G are 3D networks, optical, bio-networks and nano-networks [99]. Molecular communication is used to operate N-IoT. Nanometer-range devices can be designed by using metamaterials and graphene. BIoT is used for IoT based communication [100]. N-IoT and B-IoT are core components of emerging 6G devices. Physical layer technologies and novel routing schemes should be designed efficient biodevices and nanodevices should be developed for B-IoT and N-IoT. In addition, new models for 3D communication must be devised.

*6.3. Computing Technologies*

6G systems include various smart applications which generate large amount of data. Intelligent data analytics can be carries out by using quantum and computing technologies. In coming few years, quantum computing will pave its way to commercial market and will be a great threat to the existing cryptographic techniques. Quantum computing will revolutionize 6G network with higher data rates which is not available until now [101], [102]. It can be used in 6G to detect, mitigate and prevent from security vulnerabilities. An important characteristic of quantum communication is secure channel for data encryption. In future, quantum channel will replace noiseless classical channels to attain extreme levels of reliability. This advantage of quantum computing makes it appropriate for 6G smart applications. Similarly, integration of physical layer security scheme with post-quantum cryptography scheme will ensure secure 6G communication. Several 6G applications including terahertz communication, terrestrial wireless networks, satellite communication and underwater communication systems have potential to use quantum communication protocols e.g.

quantum key distribution (QKD). Other emerging features are quantum encryption and intelligent edge computing. These features ensure privacy and storage capability with low latency [103]. Z. Zhou et al. [104] demonstrated energy efficient edge computing for vehicular networks.

*6.4. Key Enablers*

The key enablers of 6G network are network slicing, blockchain, AI, homomorphic encryption, edge intelligence and photonics-based cognitive radio. This section discusses some key enablers for future 6G.

*6.4.1. Blockchain*

Blockchain is distributed ledger based database for secure registration and updating of transactions [105]. It aims to manage a digital ledger in a distributed and secure manner. This ledger is cryptographically sealed and all the transactions are kept in a chronological manner. It is an emerging candidate to urbanize internet services. It is an audible, decentralized and secure solution to exchange and authenticate information. Blockchain offers numerous advantages like integrity, pseudonymity, proof of provenance, non-repudiation, immutability and disintermediation. Blockchain technology is ideal for some applications due to anonymity and decentralized tamper-resistance [106]. In 2018, Jessica Rosenworcel, FCC commissioner and Mobile World Congress Americas (MWCA) focused on blockchain technology as a revolution for future generation [107]. It provides secure access for network entities and untamable distributed ledger which strengthens its security feature [108]. Blockchains are also beneficial in terms of network access and resource orchestration. X. Liang et al. [109] discussed that administration costs can be reduced through Decentralized control mechanism based on blockchain. Moreover, the spectral efficiency can also be enhanced through blockchain integration into spectrum. In 2020, F. Jameel et al. have presented a survey on reinforcement learning in blockchain and explained integration with industrial Internet-of-things (IIoT) [110]. Blockchain will enable smart health care, smart grid and smart supply chain [111-112]. It is identified as one of the key enabler to support future 6G technology. Several research efforts have been made leveraging its capability to enhance both the use cases of 6G ecosystem as well as technical aspects. Besides advantages, blockchain also faces some challenges including high energy consumption, high latency, reliability and scalability [113].

*6.4.2. Ubiquitous sensing*

Ubiquitous sensing includes 3D imaging and machine vision based video information for automatic sensing and intelligent decision making [114-115]. J. M. Segui discussed RFID tags for Ubiquitous sensing in automaker industry [116]. In future 6G, ubiquitous sensing will possibly change every avenue of life. However, it will also lead to significant problems e.g. lack of collaboration, inability to ingest and utilization of distributed information sources. It can be used in clinical diagnostics, quality control and surveillance. It has been demonstrated in clinical diagnosis and environmental monitoring. The key elements in ubiquitous sensing are implantable and wearable sensors.

*6.4.3. Homographic Encryption*

M. Salem et al. [117] used homomorphic encryption to secure biometric recognition and preserve privacy. F. Tang et al. [118] demonstrated deep learning technique for homomorphic encryption to increase security properties. Homomorphic encryption can be used to protect copyrights and preserve privacy of multimedia transmission [119]. Catak et al. proposed a novel technique to preserve privacy using homomorphic encryption and clustering methods [120]. This encryption technique is same as an arithmetic operator on an encrypted data. It offers data privacy without plain form data.

*6.4.4. Edge Intelligence*

A promising enabler for IIOT is edge intelligence as it provides smart cloud services with less cost and low latency [121]. Edge intelligence is formed by integrating edge computing and AI [122] for broader prospective as it has received a tremendous amount of attention. Edge intelligence has a wide range of applications including energy internet, smart grid, networked UAV, connected robots and autonomous driving. However, the gap lies to find out solutions for big data, coded computing, system modeling and scheduling scheme for edge intelligence. A potential challenging issue in industrial networks is to ensure edge service. Zhang et al. [121] demonstrated blockchain and edge intelligence based IIOT framework to obtain secure and flexible edge service. Edge intelligence can be implemented in cognitive internet of things to improve interactivity and sensitivity. Zhang et al. [123] introduced CIoT, a new network paradigm, to meet technical requirements such as efficient storage, generating big sensory data and integrating multiple data sources.

*6.5. Use Cases*

It is important to define new use cases for promising 6G technology. The innovative 6G services include low-latency communication, mobile broadband reliable, Nano-Internet-of-things (N-IoT), Bio-Internet-of-things (B-IoT), massive URLLC and autonomous connected vehicles. We have discussed some use cases for 6G below.

*6.5.1. Haptics communication*

It is a communication technology based on tactile sensation for human-computer interaction with computers. It is a tangible feedback system to take advantage of human's sense of touch through motion, sensation or forces. It enables physical interaction between humans and remote objects. It is an innovative research domain to understand core functions of human touch. Haptic devices like actuators and sensors allow users to sense and control objects in virtual and real world. These devices still face a gap in terms of cost effectiveness as well as degrees of freedom. This technology needs substantial design efforts to enable in 6G. Van Dan Berg et al. [124] have investigated some challenges to realize haptics communication over tactile internet. In order to realize the envisioned applications, haptics communication should offer tactile and kinesthetic control simultaneously.

*6.5.2. Holographic communication*

Holographic communication enables remote connectivity with high accuracy. Generally, it is multi-dimensional camera image communication which needs higher data rates (Tbps) [16]. Huang et al. [125] have discussed emerging trends and challenges for holographic communication in 6G.

*6.5.3. Unmanned Mobility*

This use case is related to autonomous connected vehicles which enable enhanced traffic management, smart infotainment, secure driving and unmanned mobility. Giordani et al. [126] discussed unmanned mobility with safe driving an autonomous transportation features.

*6.5.4. Bio-Internet of Things*

This technology makes use of IoT for communication of bio devices. This use case has advantage in smart health care sector. The performance characteristics of B-IoT must be defined as like N-IoT. A. Salem investigated wireless communication in THz band considering rbcs concentration in blood [65]. In 2018, S. Canovas-Carrasco et al. [66] used human hand scenario to develop nano scale communication network. Thus, B-IoT can efficiently enable 6G.

*6.6. Machine Learning Techniques*

Recently, machine learning elicited high attraction to enable wide applications. It can be a fundamental pillar for future 6G networks. Machine learning has given efficient performance in various areas including game AI, autonomous driving, language processing [127], IoT security [128], wireless-powered ambient backscatter communication [129], vehicular networks and pattern recognition. Perspectives of ML in vehicular networks in shown in figure 7. Generally, we divide ML is different categories as discussed below.

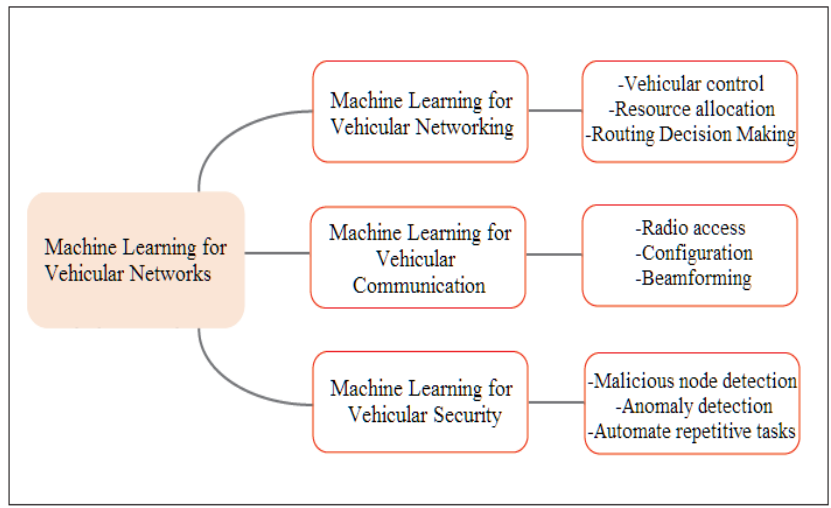

**Figure 7.** Perspectives of ML in vehicular networks

*6.6.1. Quantum machine learning*

Quantum machine learning is another most promising technology for human beings. It has emerged as an excited paradigm. Several research studies are presented in this domain [120-133]. It combines machine learning and quantum physics to design quantum machine learning models. It uses quantum devices for intelligent, accurate and fast machine learning calculations and improves control quantum systems. It is widely used in quantum mechanics and quantum biomimetic.

*6.6.2. Meta learning*

We have witnessed a dramatic rise in interested in this field of meta-learning in recent years as many studies are presented in this domain [134-136]. Meta learning has potential use in neural networks [137], speech recognition [138] and to develop curiosity algorithms [139]. Meta-learning can handle several conventional challenges of data and computation bottlenecks.

*6.6.3. Federated learning*

Federated learning has achieved widespread attention as it prevents the leakage of personal information. It has the feature to update parameters without collecting raw data. Several research studies [140-142] have focus on FL in several aspects. T. Yang et al. [143] demonstrated FL to improve Google keyboard query search. However, there are several issue e.g. security, privacy, resource allocation and cost to implement FL at large scale. FL has some inherit challenges such as incentive mechanism design, computation resource optimization and communication. Some challenges for advance machine learning based 6G are shown in figure 8.

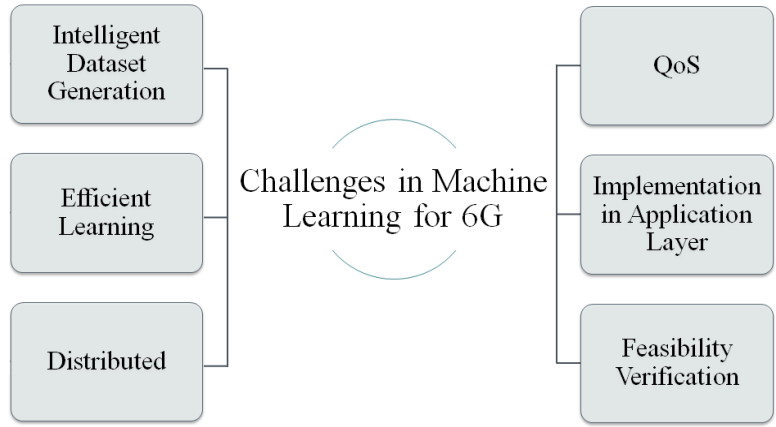

**Figure 8.** Challenges for advance machine learning based 6G.

*7. Key areas in 6G networks*

Some features of 5G have already implemented AI in various applications. However, the traditional network architecture limits AI-driven technologies. It does not support intelligent radios and distributed AI. Although realtime intelligent edge is already deployed in 5G networks but it cannot be fully controlled in realtime. However, 6G network can handle this scenario. In addition, 5G is limited to ground level; undersea and space communication is not possible. Accordingly, we have discussed some key areas and potential issues in these areas. Table VII provides a summary of these key areas.

*7.1. Real-time intelligent edge*

Implementation of Unmanned Aerial Vehicles (UAVs) networks with current technologies is not fully possible as it can only control the network with real time intelligence and extremely low latency. Although 5G technology supports self-driving, however prediction, self-awareness and self-adaption network parameters is not featured [144]. Hence, a new technology is needed to tackle these challenges. It is highly feasible by 6G technology to enable AI-assisted services. As AI will be integrated in vehicular networks, it can support numerous security algorithms. However, this integration can cause various privacy and security challenges. In [145], Tang et al. stated that both physical environment and network system should be taken into account for a vehicular network as it can mitigate malicious attacks.

*7.2. Intelligent Radio*

In previous generations, transceiver algorithms and devices were developed together. However, now transceiver algorithms and hardware can be separated. Thus, transceiver algorithm can update itself on the basis of hardware information [146]. P. Yang et al. [147] stated that software defined network techniques can enable intelligent radio signals after combining with leverage multiple high-frequency bands. Shafin et al. [148] discussed AI based cellular networks. However, several requirements must be satisfied to enable intelligent radios. Tariq et al. [25] investigated suspicious activities during communication process. While Jiang et al. [149] investigated some signal jamming problems during data transmission. There is a need to develop simple, yet highly effective security approaches as communication systems suffer from security, privacy and jamming attacks.

*7.3. Internet of Everything (IoE)*

6G networks will support Internet of Everything (IoE) which is referred as an extension of IoT including people, data, processes and things. The key idea of IoE is to incorporate different sensing devices to identify, monitor and take intelligent decisions to design new operations. The sensing devices in IoE are capable to acquire several parameters including pressure, bio-signals, light, position, velocity and temperature. These devices are utilized in different application scenarios ranging from traffic, smart cities, and digital healthcare to industrial sector. It will support intelligent decision making feature in 6G networks [146]. The incorporation of IoE and 6G will be useful to enhance the services related to body sensor networks, smart city, smart grid, connected robotics, internet of medical things and many more avenues. It is envisaged that fusion of IoE and 6G will enable various novel applications to create a new era with improved and agile features.

*7.4. 3D intercoms*

In future technology, network planning and optimization will be extended from two-dimensional to three-dimensional [114]. 6G technology will be able to feature 3D communication to support underwater, aerial and satellite communication. A 3D intercom can support this attribute with precise location and accurate time. Additionally, network resources, routing and mobility aspects also need optimization strategies in 3D intercom. By using THz band, emerging technologies like molecular and quantum communications can be used for distant communication [151]. Wei et al. [152] investigated some security attacks for authentication perspective. Similarly, performance evaluation of 6G networks in underwater environment is also unforeseen. Once 6G network operations in underwater environment are achievable, innovative applications and challenges will appear in near future. Different application scenarios empowered by 6G technologies are shown in figure 9.

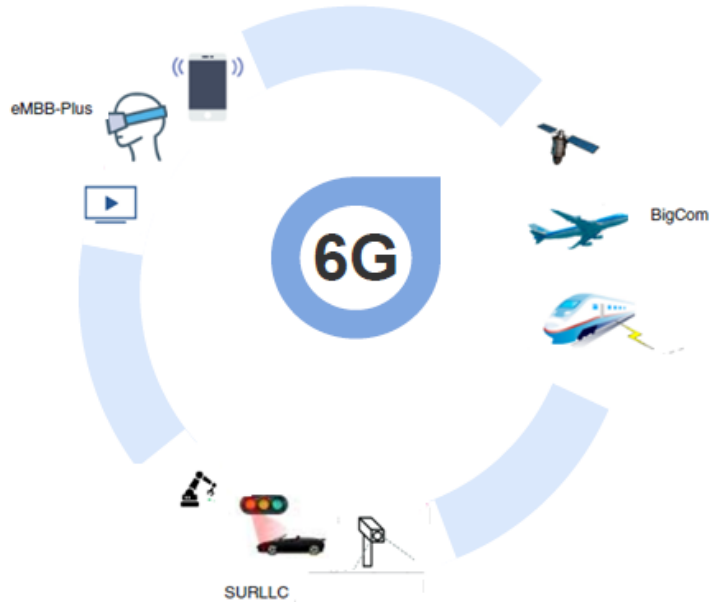

**Figure 9.** Some applications supported by 6G

TABLE VII. SUMMARY OF KEY AREAS.

| Key area | Relation to 6G | Characteristics | Summary |
|---|---|---|---|
| 3D intercoms | Coverage | Full 3D-cover | It can provide coverage at ground, space and undersea levels. |
| Intelligent radio | Communication | Self-adaptive | This framework can configure and update dynamically according to the provided hardware information. |
| Distributed artificial intelligence | Decision making capacity | Intelligent decision making | This system is capable to make intelligent decisions at various levels. |
| Real-time intelligent edge | Control capability | Real-time response | It can provide autonomous driving at an unfamiliar place in real-time. |

## 8. Vision and key features for future 6G

This section highlights various key features for future 6G networks. In this regard, Table VIII summarizes various key features such as mMTC, eMBB, eMBB-Plus, BigCom, and URLLC etc.

*8.1. Mobile Broadband Reliable Low Latency Communication (MBRLLC)*

Saad et al. proposed MBRLLC [15] by integrating eMBB and URLLC for 6G system to enable low latency and high reliability. The core aspect of MBRLLC is energy efficiency. It also considers impacts of resource utilization, rate and reliability on 6G network.

*8.2. eMBB-Plus*

eMBB-Plus [153-154] will provide high quality experience (QoE) in future 6G technology. Notably, other key features like interference and handover will be able to exploit big data. Moreover, globally compatible connection and accurate indoor positioning is also expected. There is a need to design strategic plans for eMBB-Plus without any compromise over privacy, secrecy and security of network users.

*8.3. Multi-Purpose 3CLS and Energy Services*

6G system must support multi-purpose services. It can wireless transfer power to small devices through WPT function. MPS system is good for CRAS, however, it should meet computing, control, mapping function, sensing and energy consumption performance.

*8.4. Big communications (BigCom)*

BigCom [155] in 6G will be capable to support high coverage in distinct areas. It will maintain a resource balance to establish a high data rate communication among users. Furthermore, high AI and THz band in 6G will include environmental and operational aspects for better communication.

*8.5. Human-Centric Services (HCS)*

In ref.15, authors proposed human-centric services (HCS) which require QoPE targets. Wireless BCI is a similar aspect to realize HCS in which physiology of users defines the network performance. For HCS, a function of raw QoE and QoS metrics must be defined.

*8.6. Secure ultra-reliable low-latency communications (SURLLC)*

SURLLC can be highly beneficial for vehicular communication [155-156]. SURLLC in 6G is advancement in mMTC and URLLC with high stringent demands on latency (lower than 0.1 ms) and reliability (more than 99.99%).

*8.7. Massive URLLC*

URLLC in 5G technology was introduced to meet latency for IoE applications like smart factories. Massive URLLC will keep scalability, latency and reliability into consideration. Hence, a proper framework for 6G that enables better performance for decision making, topology, architecture, reliability and delay is highly required.

*8.8. Three-dimensional integrated communications (3D-InteCom)*

There is a need to bring a radical change from 2D to 3D-InteCom model by including the high aspect of communication nodes for full dimensional MIMO architectures [156-158]. Some of the notable technologies in which 3D-InteCom can be incorporated are underwater communication, unmanned aerial vehicle (UAV) and satellite communication. Thus, a re-adjustment in 2D model which is stemmed from graph theory and stochastic geometry is required.

*8.9. Unconventional data communications (UCDC)*

Up to now, there is a lack in proper definition and meaning of UCDC [155]. However, follow facets must be discussed: human bond, tactile and holographic communication.

*8.9.1. Holographic communications*

It is expected to add glamor in 6G technology. It is a 3D technology which controls a light beam incident on any object and uses a recording device to capture resulting pattern. In real, it is insufficient to real presence scenario through 3D images without a stereo voice. In future 6G, stereo audio will be incorporated to get presence characteristics. In other words, In other words, received video or holographic data can be modified. Holographic data will use high bandwidth to transmit data over reliable network [159].

*8.9.2. Tactile communications*

Real-time conveyance or cinematic experience is possible through tactile internet [160]. Some expected advantages of this technology are interpersonal communication, cooperative self-driving and teleoperation. A haptic touch can be implemented in this technology. Realizing this technology requires some stringent needs for cross-layer architecture. It can trigger research activities to design novel physical layer schemes. It will also bring attention to design procedures e.g. protocols, handover, scheduling, queuing and buffering to meet requirements of 6G networks.

*8.9.3. Human-centric communications*

This technology will provide human access to physical features. Invariably, it will involve five human senses. A promising use case of this technology is "communication through breath" project, which makes use of exhaled breathe to read bio-profile [161].

Consequently, it will enable remote interactions with human body, biological features collection, emotion detection and disease diagnosis. Thus, to design a communication system which can realize five human senses requires interdisciplinary research efforts.

TABLE VIII. 6G SERVICES, PERFORMANCE INDICATORS AND APPLICATIONS

| Service | Performance Indicator | Applications |
|---|---|---|
| MBRLLC | Energy efficient | Autonomous drones<br>XR/AR/VR |
| eMBB-Plus | QoE | Accurate indoor positioning |
| MPS | Wireless energy transfer<br>Accurate mapping<br>Stable control | XR<br>Telemedicine<br>CRAS |
| Big communications (BigCom) | Balance resource utilization | High coverage to remote areas |
| HCS | QoPE | Efficient communication<br>Haptics |
| Secure ultra-reliable low-latency communications (SURLLC) | Low latency, High reliability | Vehicular communication |
| mURLLC | Massive reliability<br>High connectivity | Autonomous robots<br>Blockchain<br>User tracking |
| 3D-InteCom | MIMO architectures | Underwater and satellite communication |
| Unconventional data communications (UCDC) | Holographic and tactile communication | Automated driving, disease diagnosis and teleoperation |

## 9. Potential Challenges and Practical Considerations

There exist multiple challenges which can affect the performance of future 6G technology. In this section, we have explored the potential unresolved challenges of hardware design, power supply, network security, reliability, latency and user mobility. We have provided readers with motivation to address and solve some of these challenges as shown in figure 10.

*9.1. Portable and Low-latency Algorithm and Processors*

The existing artificial intelligence technologies are designed to fulfill definite requirements; however, these technologies suffer from limited migration. In this regard, a potential solution is to develop portable and low latency algorithms. Additionally, it is essential to meet accuracy and latency trade-off in these algorithms than conventional computer vision tasks. In order to provide better performance in latency critical scenarios such as medical/health and automated vehicles applications, a communication link must be established within a short interval of time. It is quite challenging to achieve low latency in few milliseconds. To attain low latency and ultra-high reliability, it is required to design powerful high-end processing units with minimum power consumption.

*9.2. Hardware Co-Design*

High density parallel computing techniques are needed in AI-assisted techniques. While certain parameters are required in wireless network architecture to enable AI-assisted communication. Furthermore, computing performance degrades in case of advance materials such as high temperature superconductors and graphene transistors. Thus, it is a key issue to miniaturize high frequency transceivers. Such as, Qualcomm and several other companies have been working to decrease size of mmWave components from meter level to smallest fingertip antennas. This issue will be more adverse for THz band. As explained in a previous study [162], optoelectronic is a promising solution which is capable to exploit the advance antennas, high-speed semiconductor and on-chip integration.

Transceiver design is a challenging issue in THz band as current designs are not sufficient to deal with THz frequency sources (>300) GHz) [163]. Current transceivers structures cannot properly operate in THz band frequencies [163]. New signal processing techniques are needed in order to mitigate propagation losses at THz spectrum. Furthermore, noise figure, high sensitivity and high power parameters must be controlled. A careful investigation of transmission distance and power is also required. Moreover, a novel transceiver design considering modulation index, phase noise, RF filters and nonlinear amplifier is needed. Nanomaterials like graphene and metal oxide semiconductor technologies can be considered to design new transceiver structures for THz devices [164]. The aforementioned metasurfaces are envisaged to support different applications involving the operation over frequencies ranging from 1 to 60 GHz. Thus, developing efficient metasurface structures which can dynamically switch the operating frequency will open a new research era to realize THZ communication.

*9.3. Power Supply*

6G has the capability to efficiently and flexibly connect autonomous mobile devices. Energy-efficient techniques become very essential in such scenarios. Currently, smartphones require novel power supply techniques for efficient performance with 6G technology. The limited battery life span of wireless devices poses a substantial design challenge. To deal with this challenge, different wireless charging methods including wireless power transfer (WPT) [165] and wireless energy harvesting have been proposed as potential solutions to offer perpetual energy replenishment in these networks. In addition, signal detection algorithms and low complexity precoding techniques can be developed for high power efficiency. On the other hand, a strategic approach to optimize WPT techniques to enable future 6G mobile devices is required to enable energy autonomy in diverse conditions. Similarly, research contributions must be dedicated to explore metasurfaces which can steer, collimate and absorb electromagnetic waves in order to utilize the main operations of metasurfaces for wireless charging of any devices over a considerable long distance.

*9.4. Network Security and Privacy Issue*

A major challenge in 6G is security and privacy problem. In 6G, integrated network security should be kept into account with physical layer security. Therefore, an intensive study is required to find new security approaches. Moreover, 5G security techniques can be extended to enable 6G. For example, secure mmWaves and massive MIMO technique can be integrated into THz band applications. H. Yao et al. [166] demonstrated a distributed key management mechanism which is a key solution for STIN. A well-integrated security mechanism can be formulated to secure privacy in 6G networks. Furthermore, an exponential growth has been witnessed in number of IoT devices in the last few years. These devices contain industrial, health care and personal IoT which can be linked to create a mesh network. 6G technology is envisaged to be the key enabler for large scale cyber mechanism within IoT scenarios. In such scenarios, distributed denial of service (DDoS) attacks will be very common as IoT devices are linked with internet. Such large-scale DDoS attacks can cause trust, privacy and security issues in the network. In future, it is important to address physical layer security (PLS) mechanism to link users to the proper source such that it can enhance the system secrecy rate. The adaptability and flexibility of PLS strategies, specifically for resource-constrained environments, together with the services provided by promising 6G technology will reveal new research directions for PLS in 6G.

*9.5. 3D Networking Reliability-Latency Fundamentals*

6G technology will support deployment of 3D applications such as 3D base stations. Research into propagation model for 3D structure is essential. Frequency utilization and 3D network planning is needed due to change in degree of freedom and altitude dimension from 2D to 3D. Furthermore, 3D evaluation metrics of rate-reliability-latency trade-off is necessary. Some recent studies [167,168] have provided brief discussions in this direction.

*9.6. Potential Healthcare Issues*

Although 6G technology can provide massive data rate at THz spectrum, but experts envisage that 6G applications are yet inchoate. THz waves propagation can effect human safety as it has three times higher photon energy level as compared to nonionizing photon [169]. International Commission on Non-Ionizing Radiation Protection [170] and Federal Communications Commission (FCC) [171] regulations are followed to reduce potential hazards. Moreover, a careful consideration is required on molecular and biological impacts of THz waves. Another promising solution to mitigate health issues is electromotive force transmission [172]. 6G will be the right approach to address the intelligent healthcare service in the future. Thus, device authentication, secure data transmission, encryption and controlling wearable devices will be a crucial security issue to be solved in 6G era. User privacy and ethical concerns of electronic health data will be major issues in future healthcare systems. There is need to develop new AI-driven models following strict ethical concerns to keep privacy and integrity aspects of healthcare records. These models must observe privacy rules and regulations implemented by the concerned authorities.

*9.7. Inteference Management*

In order to cope with the short range hindrance in wireless communication technologies, a common approach is to employ maximum of access points (APs) to enhance the link coverage in small cell scenarios. In different indoor environments, such as conference rooms and office cubicles, networks face severe interference due to a large number of access points. Interference becomes detrimental in that case where device is located closely to the interfering APs. Thus, researchers should focus on developing new interference management mechanism in order to avoid significant degradation in the performance of wireless communications technologies.

*9.8. User Mobility*

User mobility imposes a great challenge in to implement any wireless networks such as mmWaves and it severely degrades the system's performance and capacity. Therefore, it is suggested to develop adaptive, efficient and novel coding and modulation schemes to overcome channel variations. In addition, in indoor environments, which contain multiple access points (APs) to serve multiple devices, user mobility incurs rapid load fluctuations. Thus, this calls for the development of sophisticated handover mechanisms which can provide improved system's capacity, balanced load and a guaranteed QoS to realize efficient communications in future wireless networks.

*9.9. Variable Radio Resource Allocation*

For variable quality of service desiderata, a variable radio resource must be allocated to the user. It can be either variable power or bandwidth and even in some scenarios it can be both. Another challenging factor in 6G is that the signals have high penetration loss and can attenuate quickly at higher frequencies. These signals also attenuate automatically upon accessing workplaces, residences, offices and houses. As the radio waves suffer from attenuation with increasing frequency, it can face hurdles while penetrating through walls in houses and building, ultimately it affects the QoS requirements. It is therefore of high significance to design precise and stable algorithms to cope with 6G communication requirements through dynamic allocation of variable resources.

*9.10. Blockage and Shadowing control*

Sensitivity to blockage in LOS links represents a major challenge in communication technologies. Specifically, an abrupt obstruction in line-of-sight transmission between the base station and the user poses delay or even disconnection, causing a notable decrease in the system's performance and reliability. Moreover, designing a new link between another base station and user enhances the network overhead, affecting the overall network's latency. A promising solution is signal steering, which can mitigate human obstructions. However, it needs a large number of APs, which augments complexity as well as interference. Therefore, it is essential to design reliable anti-blockage mechanisms before the implementation of effective communication technologies such as mmWave communication in future 6G wireless networks.

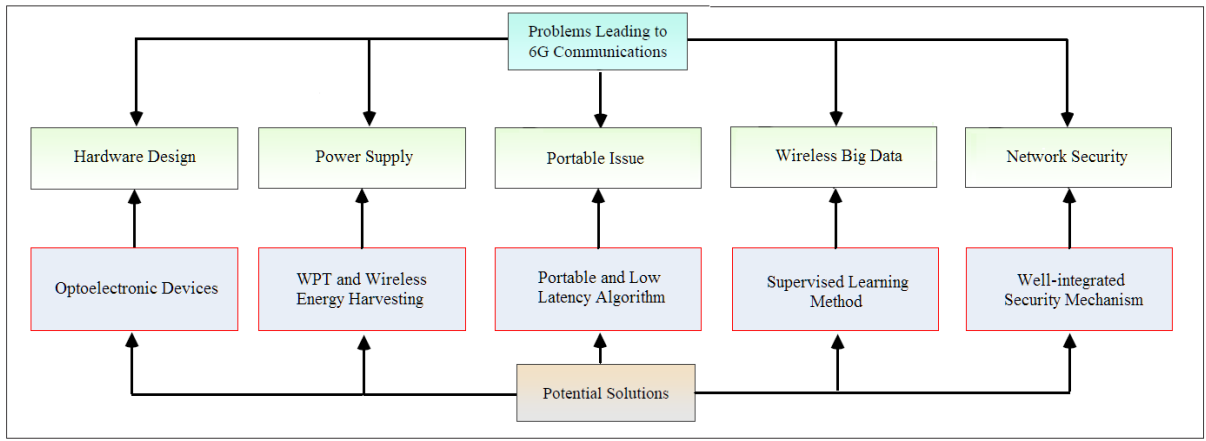

**Figure 10.** Problems in 6G and promising solutions

**10. Key Projects on 6G**

*10.1. 6G Flagship (May 2018 – April 2026)*

The 6G Flagship [58] is eight years project funded by Academy of Finland for "6G-Enabled Wireless Smart Society and Ecosystem". The aim of this project is to discover how 6G will change our lives. This project is categorized into four different research domains including devices and circuit technology, wireless connectivity, distributed computing and services and application. New 6G standards will be developed under this project for future digital societies. It has been started in cooperation with Aalto University, Oulu University of Applied Sciences, BusinessOulu and VTT technological research center of Finland. Project opportunities within 6G Flagship program include academic research, summits, symposiums, multi-partner project to tailored companies and commercialization. The academic research under this program will address communication between people, objects, devices considering privacy and security challenges. In industrial aspects, its aim is to enable a high automated and smart society. It will enable unique wireless enabled solutions for future digital societies with a tight collaboration between industrial experts from various fields. It also focuses on 5G Test Network (5GTN) providing unique possibilities to test 5G technology, components and services in real time.

*10.2. South Korea MSIT 6G research program*

The government of South Korea aims to initiate a 6G pilot project in 2026. 6G services in South Korea will be commercially available between 2028 and 2030 [173]. The government expects to invest $169 million between 2021 to 2026 in order to enable basic 6G technology. The government's strategic plan for 6G is based on preemptive development of 6G technology, new standards, high-value-added patents, research and development (R&D) and industrial collaborations. The initial strategic tasks include hyper-trust, hyper-intelligence, hyper-space, hyper-precision, hyper-bandwidth and hyper-performance. Major research areas which have been adapted for 6G pilot project include smart factories, smart cities, self-driving cars and digital healthcare immersive content. The South Korean Ministry of Science and ICT (MSIT) has also formulated the "6G R&D Strategy Committee" [173] which contains public universities in South Korea, government agencies and small/large scale device manufacturers to manage 6G related projects. The goals of this 6G pilot project are: 1) to use AI within entire network, 2) to extend connectivity up to 6.2 miles

from the ground, 3) to reduce latency up to 0.1 ms, 4) to achieve 1Tbps data rate and 5) to enable various security features to secure entire network.

*10.3. Japan B5G/6G Promotion Strategy*

The Japanese government will earmark $482 million (50 billion yen) to promote R&D initiatives under 6G promotion strategy. This fund is allocated to support 6G test-bed facility for institutional and industrial testing of its designed technologies. Japanese government plans to use 30 billion yen from this fund in coming years to support R&D in 6G technology. The government also plans to use 20 billion yen to design a facility to be used by companies and other collaboration partners to test their developed technologies. Japan envisages designing and showcasing core technologies in 2025 while 6G will be commercially launched around 2030 [174]. The 6G vision includes scalability, autonomy, reliability, ultra-security and resiliency, ultra-low latency, ultra-fast and large capacity, ultra-numerous connectivity and ultra-low power consumption [174].

*10.4. INSPIRE-5Gplus*

INSPIRE-5Gplus, Research and Innovation (RIA) project under EC H2020, is a 36 months project started in 2019 [175]. It has various project partners such as Universidad de Murcia, National Centre for Scientific Research Demokritos and TAGS etc. [175]. INSPIRE-5Gplus is completely devoted to strengthen security of 5G and B5G networks considering different features including learning models, use cases, architecture, novel enablers and network management. It is based on two approaches: 1) by leveraging existing assets and 2) by introducing novel solutions through blockchain, AI and ML. This project will address key security challenges for efficient and concrete realization of 5G. The outcomes of this project will serve the crucial objectives of pervasive trust and intelligent security. It will also deliver unique assets to enable trusted and intelligent multi-tenancy i.e. liable, evidence-based, and confident across holistic architecture of multi-tenants network.

*10.5. AI@EDGE*

The key objective of AI@EDGE project is to design a secure AI-assisted platform for edge computing in B5G networks [176]. It will enable frameworks to create, utilize and adapt trustworthy, reusable and secure AI/ML models. The aim of this project is to design a connect-computer fabric in order to create and manage secure, elastic and resilient end-to-end slices. These slices will support an extensive range of AI-enabled applications Moreover, trusted networking and privacy preserving ML techniques will be adapted to ensure privacy and framework usage without disclosing sensitive information. This project will focus on breakthroughs such as multi-connectivity, provision of AI-enabled application, privacy preserving, AI/ML for closed loop automation and ML for multi-stakeholder environments. The AI@EDGE platform will be performed through four high impact use cases including smart data and content curation for in-flight entertainment services, edge AI aided monitoring through UAVs in BVLOS operation, resilient and secure orchestration of large IoT networks and virtual validation of cooperative vehicular networks [176].

*10.6. Hexa-X (January 2021 – June 2023)*

The Hexa-X project [177] is initiated with the vision to firmly anchor human and digital worlds through a fusion of 6G key enablers. The vision of Hexa-X demands an x-enabler fabric of trustworthiness, extreme experience, global service coverage, sustainability, networks of networks, operational resilience, integrity of secure communication and connected intelligence. This project aims to investigate new key enablers in 6G for

- Connected intelligence via AI-driven air interface
- High resolution localization and sensing
- Management of future networks
- Radio access technologies at higher frequencies
- 6G architectural elements for dynamic dependability in network

Considering above aspects, Hexa-X project has been started under 6G flagship to bring together the main industrial stakeholders, network operators, network vendors as well as the academia researchers from most prestigious European research centers to bring an integrated contribution in research and development (R&D) towards 6G.

*10.7. 5GZORRO*

5GZORRO is also an EC H2020 RIA project which aims to investigate new set of solutions to enable zero-touch privacy, security and trust in network and security management in distributed multi-stakeholder environments [178]. It will enable smart contracts for dynamic spectrum allocation, ubiquitous connectivity and will support required agility. It will design architecture for 5G network in a trusted and secure way. The target stakeholders of 5FZORRO are regulators, spectrum owners, virtual slice operators, telecom services providers and active/passive facility owners.

*10.8. NEW-6G and RISE-6G*

Recently, two new European initiatives have been announced as NEW-6G and RISE-6G [179]. NEW-6G refers to Nano Electronic and Wireless for 6G. RISE-6G is launched under 5G PPP focused on reconfigurable intelligent surfaces (RIS). Both the projects will be led by Atomic Energy Commission and French Alternative Energies. E.U. has allocated €6.49 million for RISE-6G under Horizon 2020 (H2020) program. It will enable ubiquitous wireless connectivity, ultra-massive, instantaneous, data-driven as well as connected intelligence according to an article published in November 2020 including RISE-6G principal investigator Marco Di Renzo. RISE-6G will perform preliminary tests in real-time scenarios such as train station. RISE-6G will help to investigate a broader range of subjects: deployment, infrastructure, network optimization, innovative technologies and fundamental science. Furthermore, NEW-6G will support unprecedented opportunities to rethink the role of nano-electronics and to promote innovative ideas, share knowledge, encourage cooperation and establish roadmaps [179].

*10.9. ATIS' Next G Alliance*

Several western companies including QUALCOMM, Nokia and AT&T have initiated the Next G Alliance through a U.S. based standards organization named the Alliance for Telecommunications Industry Solutions (ATIS) [179]. ATIS initiated this project to lay out the foundation of 6G for a vibrant marketplace for services and products in North America. The coalition already announced a team devoted to produce a 6G roadmap for the next decade to become a strong global mobile technology leadership. This group has 43 founders including some tech giants like Facebook, Apple, Google and Microsoft etc. Unlike other programs which foster 6G, the Next G Alliance started from a private sector-led initiative whose objective is to influence U.S. funding agencies which will incentivize the industry [180]. Besides funding and research, the Next G Alliance aims to encompass a high-level strategic perspective of standards and developments, manufacturing standardizations and market readiness. The main idea to bring collaboration between diverse segments of government, research institutes and industry, together with a strong emphasis on technology commercialization and engaging international community into discussion about standardizations.

*10.10. Other Projects*

Several projects have been launched under Horizon 2020 (H2020) program. 6G BRAINS [181] has been launched to bring AI-driven DRL to perform resource allocation with new spectrums including optical wireless communication (OWC) and THz to improve the performance regarding latency, reliability and capacity for future industrial networks. Similarly, DEDICATE 6G [182] has been launched with this vision to transform B5G networks into a smart connectivity platform which will be resilient, ultra-fast and highly adaptive to support human centric services. It will address trust, privacy and security assurance for novel interaction between digital systems and humans. Additionally, MARSAL [183] has been initiated with this aim to develop an entire framework for orchestration of network resources in B5G by using optical wireless infrastructure and radically enhancing the flexibility of this architecture. Its objective is to enable such a mechanism which offers security and privacy to application data and workload. Furthermore, DAEMON [184] aims to set forth a pragmatic strategy for network intelligence design. The main objectives of DAEMON include extremely high reliability, reduced energy footprint of mobile network and extremely high performance in real-time scenarios. Simultaneously, REINDEER [185] aims to design smart connectivity technologies with uninterrupted availability, perceived zero latency and resilient interaction experiences. It will develop a novel wireless access infrastructure called RadioWeaves as a massive distributed antenna array composed of fabric of distributed radio, computing and storage components. Its objective is to design algorithms and protocols to enable new resilient interactive services which require real-space and real-time cooperation for future intuitive care, immersive entertainment and robotized industrial environments.

**11. Potential Applications of 6G**

Every new wireless generation introduces some novel applications. Here, we have discussed several potential applications for future 6G wireless networks. Table IX provides a summary of these 6G applications.

*11.1. Multi-sensory XR applications*

The advantages of 5G technology such as high bandwidth and low latency have extended the VR/AR experience for 5G network users. However, there are several potential issues which must be addressed in future 6G network in order to enhance this VR/AR experience. Several sensing devices can be deployed to collect sensory data. Hence, a new feature extended reality (XR) can be realized from eMBB and URLLC. Extended reality (XR) is an appealing technology which contains Augmented Reality (AR), Virtual Reality (VR) and Mixed Reality (MR). 6G will support the advancement of XR in various use cases including robot control, healthcare, video conferencing, entertainment and virtual tourism. This requires extreme low latency, high resolution, extreme data rates and strong connectivity, which is envisioned to be supported by 6G. Additionally, several aspects including devices diversity, low overhead and high scalability should be taken into account while designing the security mechanism of XR. The major security concerns are malicious threats. Access control, encryption, authentication and internal communication. In [186], R. Chen et al. [186] briefly discussed security challenges in URLLC services. While in [187], J. M. Hamamreh et al. [187] proposed an approach to enhance security against malicious attacks in URLLC. Furthermore, authors in [188] have suggested a 3D model which addresses secrecy threats in XR applications.

*11.2. Connected robotics and autonomous systems*

Academia researchers and industrial experts have shown considerable interest in future transport systems such as internet of vehicles, cooperative vehicular networks, intelligent robotics and self-driving. Almost 50 leading technological and automotive companies have shown interest to invest in autonomous vehicle technology. In future, connected autonomous vehicles (CAV) technologies will introduce a new service ecosystem such as self-driving public transports. Specifically, AI-enabled future vehicular networks will pave the way towards intelligent transport system (ITS). In [189], Strianti et al. discussed automatic handling, caching and resource control in network. They designed a complete automated factory based on UAVs, database and cloud services. Similarly, UAV network, new algorithms and advanced hardware can be implemented in different operations including agriculture, emergency, construction and fire control. In future, fully automated vehicles and robots will participate in maintenance process, monitoring, operation and real-time diagnostics. Intelligent robots will be deployed at harsh environments for communication and research tasks. Highly reliable and self-organized features of automation will bring a revolution is several aspects of daily life. Such innovations will pave a way to develop new cities that are smart, greener, sustainable and productive.

*11.3. Wireless brain-computer interactions*

The key idea behind wireless BCI is to connect human brain with any device. This device can be located inside or outside the human body. One potential feature of wireless BCI is to support disabled people by controlling auxiliary equipment. It is envisaged that wireless BCI will become an integral part of future 6G technology. In [190], Chen et al. proposed a BCI mechanism to accelerate spelling. Besides its advantages, BCI system faces several security threats such as encryption and malicious attacks. To tackle these security challenges, authors in [191-192] have briefly highlighted security issues, hacking applications and prevention methods to overcome these security challenges.

*11.4. Accurate indoor positioning*

Global positioning system (GPS) has shown significant role in outdoor environments. However, indoor position systems still require research focus to overcome complicated indoor EM propagation. Several studies are presented over indoor positioning system [193-196]. New functionalities of full-fledge services are envisaged with accurate and reliable indoor positioning systems. It is possible to realize these services in future 6G technology.

*11.5. Intelligent Internet of medical Things (IIoMT)*

It is envisaged that 6G will bring revolution in healthcare sector. In future, 6G will overcome space and time barriers to perform medical tasks beyond boundaries. Intelligent vehicles will enable Hospital-to-Home (H2H) service. Diverse intelligent sensors and wearable devices will assist to detect real-time accident and automatic surgery. IIoMT will remove space and time barriers. High speed communication based telesurgery will be utilized by remote doctors to perform surgery.

The doctors will operate telesurgery through tele-assist, verbal or telestration [197]. For verbal, doctors will use holographic communication to obtain better visual of surgery. They can tele-assist the surgical operations through haptic or tactile communication. For telestration, they will use VR and AR. An overview of telesurgery is presented in figure 11. In 2019, China has already made a remarkable feat by performing 5G remote brain surgery. With the help from Chinese technology giant Huawei and China Mobile, China's PLA General Hospital (PLAGH) successfully performed the operation through 5G technology where doctor was 3000 km away from patient [198].

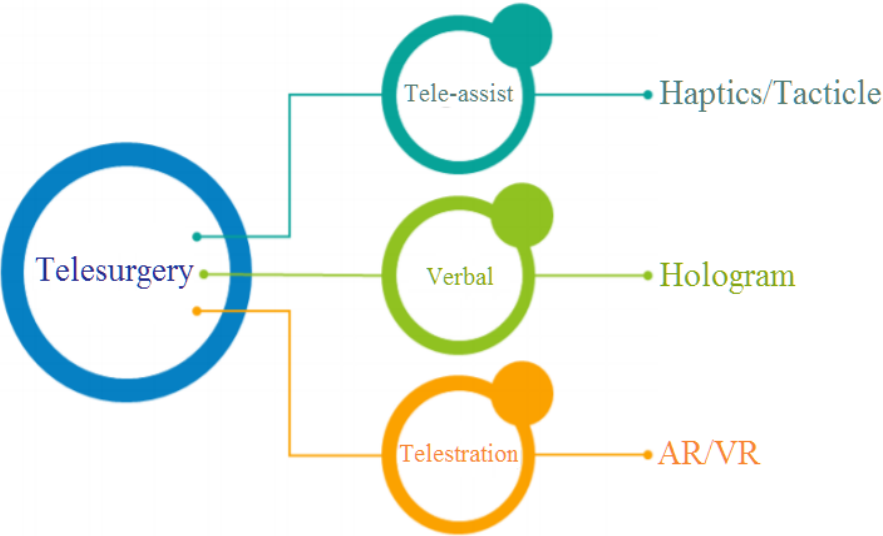

**Figure 11.** An overview of telesurgery

*11.6. Internet of Nano Things (IoNT)*

Nanotechnology has given excellent opportunities to design advanced material based nanodevices for medical and industrial use [197]. Nano-things have the ability to perform basic functionalities of sensing and actuation at a high speed while having have low data storage capacity. Generally, the idea of IoNT is derived by merging nanotechnology with IoT. In IoNT, nanosensors are connected through a nanoscale network to exchange data. Nanosensors or nano-things can communicate over a short distance by using Internet of Nano Things (IoNT) [199]. Typical architecture of IoNT is presented in figure 12.

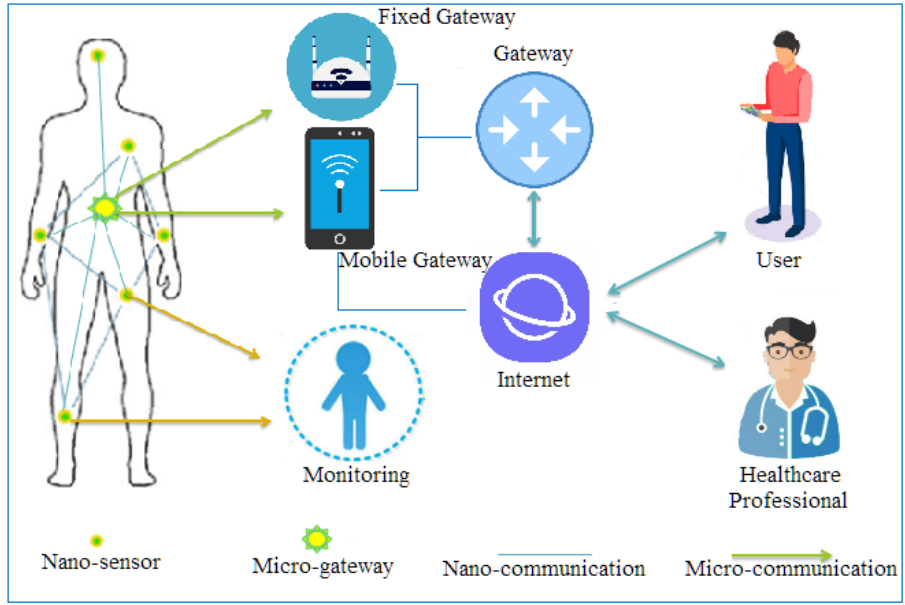

**Figure 12.** Typical architecture of IoNT

IoNT based communication can be implemented via THz or molecular communication. THz communication is more speedy, reliable and secure rather than molecular communication [200]. Future 6G technology with >1 Tbps speed will enable IoNT with a smooth data transmission. Furthermore, it will be easy to control IoNT with massive number of nano things with high density 6G technology. IoNT is expected to bring remarkable revolution in modern healthcare [201]. IoNT deployment is also complemented by other associated technologies as shown in figure 13.

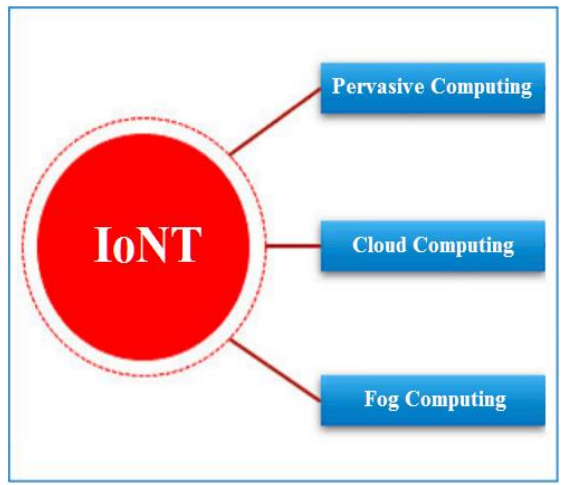

**Figure 13.** IoNT and allied technologies

TABLE IX. SECURITY, PRIVACY AND CHALLENGING ISSUES IN 6G APPLICATIONS

| Reference | Application | Security, Privacy and Challenging Issue |
|---|---|---|
| [186] | Multi-sensory XR applications | Communication |
| [191] | Wireless brain-computer interactions | Malicious Attack |
| [193] | Accurate indoor positioning | Multi-access |
| [195] | Accurate indoor positioning | Positioning |

| [201] | IoNT | Limited memory space and computational capability |
|---|---|---|
| [202] | IIoMT | QoL |
| [203] | Multi-sensory XR applications | Access control |
| [204] | Wireless brain-computer interactions | Encryption |
| [205] | Connected robotics and autonomous systems | Authentication |
| [206] | Connected robotics and autonomous systems | Communication |

*11.7. Edge Computing for Consumer Electronics (ECCE)*

The edge computing characteristic of 5G enables research fraternity and industrial experts to reconsider innovative use-cases to realize an extensive range of applications. The future wireless technologies such as B5G or 6G are envisaged to efficiently support low-latency and high-capacity short-range applications. In this regards, it is expected that future consumer electronics (CE) will effectively support wireless capabilities of B5G/6G. Nawaz et al. [207] proposed this concept of ECCE to facilitate required computing services to consumer electronics (CE) by considering e-URLLC wireless connectivity as shown in figure 14. Several CE devices can be seen in proposed ECCE framework to support eHealth, surveillance, virtual reality and entertainment etc. The proposed concept is expected to bring evolution in latency, reliability and link-speed to perform tasks locally at the devices in the B5G/6G era. The anticipated innovations contain: 1) processor-less devices, 2) inter-chip communication through THz links and 3) removing cabling requirements between processor and associated user interface.

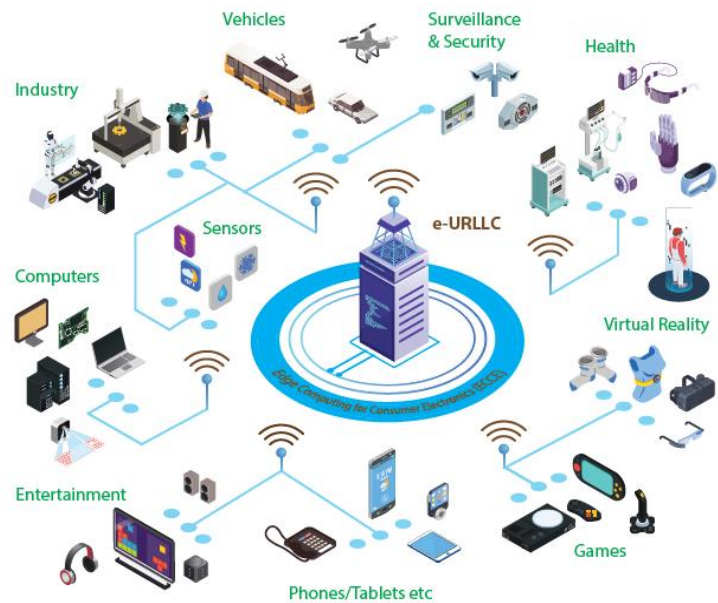

**Figure 14.** Edge computing for consumer electronics (ECCE) [207]

## 12. Conclusion

During the global deployment of 5G, both academicians and industrial experts have started realizing 6G with the aim to strengthen the competitive advantages of future wireless technologies. To support this vision, we have highlighted most promising research lines from recent literature. The future 6G technology will focus to establish communication links among objects, devices, users and industries. Performance analysis of network transmission is no longer only paramount parameter; AI, IoT and blockchain have become essential candidates. It is expected that 6G technology will keep penetrating into ubiquitous spaces, human-perceived actions and virtual societies. It will offer intelligent, deep, reliable, secure, seamless and holographic network architecture. The main contributions involve several industrial projects and research activities around the globe to support the vision of 6G. Furthermore, 6G will support several promising technologies including holographic communication,, tactile communication and visible light communication. In future, B5G/6G technologies will enable smart services and faster technologies than the existing technologies. In this concern, the existing security approaches for 4G/5G will not be sufficient to protect future 6G network. Thus, the basic parameters, such as authenticity, availability, integrity and confidentiality must be addressed in the future 6G network. Similarly, privacy-by-design must be incorporated to meet the demands of user, identity, location and data privacy. In summary the research fraternity must think to develop innovative privacy and security solutions with low-cost, ease integration and high security. This review article starts by providing the historical overview of wireless generations and associated pivotal elements to foster future 6G network. Then, we profoundly examined ongoing research progress, technological breakdown, potential issues associated with future 6G technology. This paper also outlines the key technologies, use cases and key enablers of 6G networks along with providing a prospective on future aspects. Finally, we conclude this article by shedding some light over key projects and potential applications of future 6G wireless network. We believe this review will open new horizons for future research directions by accelerating the interest of the research community towards future wireless networks innovations.

**Conflicts of Interest:** The author declares no conflict of interest.